\documentclass[1p,number,sort]{elsarticle_arXiv} % oder: 1p, 5p und twocolumn
\usepackage{etoolbox}
\patchcmd{\abstract}{Abstract}{\vspace{-8mm} \begin{center}\textnormal{{\scriptsize The published version of this article is available in \href{http://dx.doi.org/10.1098/rspa.2015.0388}{Proceedings of the Royal Society~A}.}}\\[2ex]\end{center}Abstract}{}{}

\usepackage{pifont,geometry,graphicx}
\usepackage{natbib}
\usepackage[hidelinks]{hyperref}
\usepackage{marginnote}
\usepackage{amsmath,amsthm,amssymb,amsfonts}
\usepackage{gnuplottex}
\usepackage{graphicx}
\usepackage{epsfig}
\usepackage[usenames,dvipsnames]{xcolor}
\usepackage{fancyvrb}
\usepackage{stmaryrd}
\usepackage{upgreek}
\usepackage{caption}
\usepackage{subfigure}
\usepackage{gensymb}
\usepackage{hyperref}
\hypersetup{
    colorlinks,
    linkcolor={blue!30!black},
    citecolor={blue!50!black},
    urlcolor={blue!80!black}
}
\renewcommand{\d}{{\,\rm  d}}

\renewcommand{\div}[1]{{\rm div }\left( #1 \right)}

\newcommand{\fsym}[1]{{\rm sym }( #1 )}

\newcommand{\fempty}[1]{{}}
\newcommand{\f}[1]{\mbox{$ #1 $}}
\newcommand{\topic}[1]{\\[0.8ex]{\bf  #1.}}

\newcommand{\sty}[1]{\mbox{\boldmath $#1$}}
\newcommand{\styy}[1]{{\mathbb{#1}}}
\newcommand{\fa}{\sty{ a}}

\newcommand{\fd}{\sty{ d}}
\newcommand{\fe}{\sty{ e}}

\newcommand{\fn}{\sty{ n}}

\newcommand{\ft}{\sty{ t}}
\newcommand{\fu}{\sty{ u}}

\newcommand{\fx}{\sty{ x}}

\newcommand{\fzero}{\sty{ 0}}
\newcommand{\fA}{\sty{ A}}
\newcommand{\fB}{\sty{ B}}

\newcommand{\fM}{\sty{ M}}

\newcommand{\ffC}{\styy{ C}}

\newcommand{\fsigma}{\mbox{\boldmath $\sigma$}}

\newcommand{\fxi}{\mbox{\boldmath $\xi $}}

\newcommand{\feps}{\mbox{\boldmath $\varepsilon $}}

\newcommand{\cB}{{\cal B}}

\newcommand{\cD}{{\cal D}}

\newcommand{\cP}{{\cal P}}

\newcommand{\figref}[1]{Fig.~\ref{#1}}
\newcommand{\eqreff}[1]{Eq.~\eqref{#1}}

\newcommand{\equreff}[2]{Eqs.~\eqref{#1}~and~\eqref{#2}}

\newcommand{\secref}[1]{Section~\ref{#1}}

\newcommand{\tabref}[1]{Table~\ref{#1}}

\newcommand{\cip}[1]{\citep{#1}}
\newcommand{\cit}[1]{\citet{#1}}
\newcommand{\blds}[1]{\mbox{\scriptsize \boldmath $#1$}}
\newcommand{\microm}[0]{\f{\upmu}m}
\newcommand{\EE}{\begin{equation}}
\newcommand{\Ee}{\end{equation}}
\newcommand{\FF}[1]{\begin{figure}[#1]\CT}
\newcommand{\Ff}{\end{figure}}
\newcommand{\FC}[1]{\caption{#1.}}
\newcommand{\TT}[1]{\begin{table}[#1]}
\newcommand{\Tt}{\end{table}}
\newcommand{\TC}[1]{\caption{#1.}}
\newcommand{\TB}[1]{\begin{tabular}{#1}}
\newcommand{\Tb}{\end{tabular}}
\newcommand{\II}[1]{\begin{itemize}[#1]}
\newcommand{\Ii}{\end{itemize}}
\newcommand{\EN}[1]{\begin{enumerate}[#1]}
\newcommand{\En}{\end{enumerate}}
\newcommand{\ii}{\item}
\newcommand{\CT}{\centering}
\newcommand{\gaq}{\gamma_{\rm eq}}
\newcommand{\dgaq}{\dot\gamma_{\rm eq}}
\newcommand{\caq}{{\roneC{\zeta}}}
\newcommand{\dcaq}{{\roneC{\dot\zeta}}}
\newcommand{\orange}{\color{black}}

\newcommand{\blue}{\color{black}}
\newcommand{\red}{\color{black}}
\newcommand{\green}{\color{black}}

\newcommand{\redsw}{\color{black}{}}
\newcommand{\blueeb}{\color{black}{}}
\newcommand{\ortb}{\color{black}{}}
\newcommand{\oreb}{\color{black}{}}
\newcommand{\revC}[1]{{\color{black}{#1}}}
\newcommand{\roneC}[1]{{\color{black}{#1}}}
\newcommand{\rtwoC}[1]{{\color{black}{#1}}}
\newcommand{\rthrC}[1]{{\color{black}{#1}}}
\newcommand{\rrevC}[1]{{\color{black}{#1}}}
\newcommand{\rroneC}[1]{{\color{black}{#1}}}
\newcommand{\rrtwoC}[1]{{\color{black}{#1}}}
\newcommand{\rrthrC}[1]{{\color{black}{#1}}}
\newcommand{\rev}{\color{black}}
\newcommand{\rone}{\color{black}}
\newcommand{\rtwo}{\color{black}}
\newcommand{\rthr}{\color{black}}
\makeatletter
\let\@afterindenttrue\@afterindentfalse
\makeatother
\begin{document}
%% TITLE DETAILS %%
 \title{Equivalent Plastic Strain Gradient Plasticity with Grain Boundary Hardening and Comparison to Discrete Dislocation Dynamics}%\tnoteref{t1,t2}}

%% AUTHOR DETAILS
 \ead[url]{http://www.itm.kit.edu/cm/, http://www.iam.kit.edu/cms/}
 \author[focal,rvt]{E.~Bayerschen\corref{cor1}}%\fnref{fn1}}
 \ead{eric.bayerschen@kit.edu}

 \author[izbs,rvt]{M.~Stricker}
 \ead{markus.stricker@kit.edu}

 \author[rwth]{S.~Wulfinghoff}%\fnref{fn1}}
 \ead{stephan.wulfinghoff@rwth-aachen.de}

 \author[izbs,rvt]{D.~Weygand}
 \ead{daniel.weygand@kit.edu}

 \author[focal,rvt]{T.~B\"ohlke\corref{cor2}}%\fnref{fn1}}
 \ead{thomas.boehlke@kit.edu}

 \cortext[cor1]{Corresponding author}
 \cortext[cor2]{Principal corresponding author}

 \address[focal]{Institute of Engineering Mechanics (ITM), Chair for Continuum Mechanics}
 \address[izbs]{Institute for Applied Materials (IAM)}
 \address[rvt]{Karlsruhe Institute of Technology (KIT), Kaiserstr.\ 12, D-76131 Karlsruhe, Germany}
 \address[rwth]{Institute of Applied Mechanics, RWTH Aachen University, Mies-van-der-Rohe-Str. 1, D-52074 Aachen, Germany\vspace{-5mm}}
%% ABSTRACT DETAILS DW

\begin{abstract}
The gradient crystal plasticity framework of \citet{wulfinghoff2013gradient}, {\rev incorporating} an equivalent plastic strain~\f{\gaq} and {\rev grain boundary yielding}, is extended with grain boundary hardening. By comparison to averaged results from many discrete dislocation dynamics (DDD) simulations of an aluminum type tricrystal under tensile loading, the new hardening parameter \rrevC{of} the {\rev continuum} model is calibrated. Although the grain boundaries (GBs) in the discrete simulations are impenetrable, {\rev an infinite GB yield strength, corresponding to} microhard GB conditions, is not applicable in the continuum model. A combination of a finite GB yield strength with an isotropic bulk Voce hardening relation alone also fails to {\rev model the} plastic strain {\rev profiles} obtained by DDD. Instead, a finite GB yield strength in combination with GB hardening depending on the equivalent plastic strain at the GBs is shown to give a better agreement to DDD results. The differences in the plastic strain {\rev profiles} obtained in DDD simulations by {\rev using different orientations of} the central grain could not be captured. This indicates that the {\rev misorientation dependent} elastic interaction of dislocations reaching over the GBs should be included in the continuum model, too.
\end{abstract}
% % % % %  \begin{abstract}
% % % % %   In this work the strain gradient crystal plasticity framework of \cite{wulfinghoff2013gradient}, considering an equivalent plastic strain~\f{\gaq} and a finite grain boundary yield strength, is extended with additional grain boundary hardening. By comparison to discrete dislocation dynamics simulations of an aluminum type tricrystal under tensile loading the new hardening parameter in the gradient model is calibrated. It is shown that although the grain boundaries in the discrete simulations are impenetrable, microhard grain boundary conditions are not applicable in the continuum model and instead a finite grain boundary yield strength in combination with grain boundary hardening has to be used to reproduce the pile-ups of dislocations observed in the discrete simulations. A combination of the finite grain boundary yield strength with an isotropic bulk Voce hardening relation alone, does not lead to coinciding plastic strain profiles of continuum and discrete simulations. With the hardening based on the equivalent plastic strain of the grain boundaries, instead, the overall mechanical response and the local response, i.e., averaged plastic strain profiles from the discrete simulations, can be reproduced with the gradient model. Thereby, the value of discrete simulations as a source of data for gradient continuum modeling is demonstrated.
% % % % %  \end{abstract}
 \begin{keyword}
  Grain boundary hardening, grain boundary yielding, strain gradient plasticity, \rroneC{discrete dislocation dynamics}.
 \end{keyword}
%% MAKE THE TITLE
\maketitle
\section{Introduction}
\rroneC{Microstructural characteristics influence the material behavior of multicrystalline metals. Their mechanical response is controlled by the dislocation microstructure, their interactions, and their complex interplay with other microstructural characteristics like, e.g., precipitations and grain boundaries (GBs) \cip{ashby1970deformation}. Dislocations can, e.g., pile up at GBs, be absorbed into the GBs, or be transmitted into the adjacent grains \cip{lee1990tem}. To account for these GB mechanisms in continuum models is still an open challenge, and only a limited number of three-dimensional models exist (e.g., \cip{gurtin2008theory}). In the present work, the focus is on GB effects on the local material behavior. On that account, a phenomenological model for GB hardening is incorporated into a gradient plasticity (GP) framework. Regarding its prediction of the plastic strain close to GBs, it is evaluated by comparison to discrete dislocation dynamics (DDD) results.}\\
\rrevC{Gradient plasticity theories are commonly used to take into account microstructural characteristics in continuum modeling.} In order to model, e.g., size effects \cite{fleck1994strain}, an internal length scale is introduced. Therefore, an additional (defect energy) contribution to the Helmholtz free energy of the bulk is usually considered. This contribution is formulated, for example, using the gradients of plastic slip or the plastic deformation gradient (e.g., \cip{gurtin2000plasticity}). Such an extension requires equilibrium conditions for the additionally arising microstresses. Thus, higher order boundary conditions (BCs) are needed, as well. \rrtwoC{Choices for these can, but not necessarily must, result from the requirement of achieving a null-expenditure of power by the microstresses. \rroneC{In~\cip{gurtin2005boundary}}, two such choices are discussed. These give upper and lower bounds for the defect transfer restrictions imposed by the boundary / grain boundary tractions. They are referred to as ``microscopically free'' (microfree) and ``microscopically hard'' (microhard) conditions.} \rrtwoC{The microfree case corresponds to vanishing tractions for the microstresses, i.e., the tractions do not impede the flow of dislocations which itself is modeled, e.g., by plastic strain rates. An unrestricted flow of dislocations across the boundary corresponds to continuous plastic strain rates. Whether these are continuous also depends on the material model applied and the mechanical boundary conditions.} For the opposing microhard condition, the slip rates are set to vanish on the boundary. This condition corresponds to a non-passing restriction on dislocations. Both BCs can also be applied for the case of only one accumulated plastic slip. In~\cip{gurtin2010mechanics}, this is discussed, e.g., for the gradient theory of Aifantis~\cip{aifantis1984microstructural,aifantis1987physics}. However, these two bounds for BCs are not applicable to all types of GB behavior.{\rev\ \cit{van2013grain}, e.g., give an overview about the possible interactions between dislocations and GBs. \rtwoC{One approach to achieve GB behavior inbetween the above described ``simple'' conditions is the introduction of a GB energy (see, e.g., the quadratic energies in \cip{gudmundson2004unified,aifantis2005role}, and the more general forms in \cip{fredriksson2007modelling,voyiadjis2014theory}). Grain boundary hardening can be modeled with such energy approaches (see also {\cip{aifantis2006interfaces,voyiadjis2009formulation}}).}}\\
\roneC{The consideration of GBs in gradient plasticity models requires to specify the continuity of quantities like plastic strain and microstresses {at these interfaces}. In \cit{aifantis2006interfaces}, continuity of the plastic strain is assumed. Jumps in its gradient, however, are allowed for. This model has shown good agreement with \rrevC{experiments} for a one-dimensional case. In \cit{gurtin2005boundary}, jumps in the Burgers vector flow across the GB are considered. The higher-order stresses are, however, assumed continuous across the GB. For a detailed overview on modeling approaches with regard to the continuity of higher-order quantities see \cip{voyiadjis2009formulation}. In the field theory of defects by \cit{fressengeas2012tangential}, the tangential continuity of different elastic/plastic tensors across an interface is derived from the conservation of Burgers and Frank vectors, and compatibility conditions are established at multiple junctions.}\\
\rtwoC{Naturally, continuum models and the employed hardening relations are intended to be calibrated to experiments. \rrtwoC{Latent hardening models (e.g., \cip{kocks1966latent,lavrentev1980type,franciosi1982multislip}) are commonly used (e.g., \cip{asaro1985overview,anand1996computational}). They take hardening occurring in other (than primary) slip systems due to slip in primary slip systems into account. The Mises-Hill framework of \cit{gurtin2014gradient} considers self hardening \rrtwoC{(i.e., slip systems harden due to their own plastic slip) and latent hardening}.} A hardening rule is introduced on the individual slip systems and takes accumulated plastic slips into account.} Furthermore, hardening approaches based on scalar plastic strain measures have been developed (see \cip{bouvier2005modelling} for a brief overview). Such approaches allow for rapid parameter calibration with a manageable number of experiments.\\
However, it is challenging to develop appropriate experimental settings that can be easily varied to investigate several test cases. To overcome these experimental challenges, DDD simulations are used in the \rrevC{present} work. The physically detailed modeling of plasticity by DDD results in information about the effects of the dislocation interactions. {\rev In such approaches (e.g., \cip{Csikor2007,Senger2008,Liu2011,Srivastava2013,Quek2014,Imrich2014}), the plastic response is directly computed from individual dislocation motions. Dislocations are described as elastically interacting line defects, and are discretized as polygons. Physical mechanisms of dislocation glide, cross-slip and reactions are treated with constitutive rules. In the DDD framework of the present work (\cit{Weygand2002,Weygand2009}, \cit{Siska2009}), GBs confine dislocation motion \rrevC{to remain} within the respective grains. Nevertheless, total transparency {\rev of the GBs to} stress and displacement fields of each dislocation (and, thus, the elastic interaction of dislocations across GBs) is preserved.} From the results of DDD simulations, conclusions can be drawn to incorporate the dislocation behavior in the GP model phenomenologically. \cit{bardella2013latent}, e.g., use a strain gradient extension of conventional latent hardening and benchmark it with DDD results for a simple shear problem. In \cit{aifantis2009discrete}, a tricrystal tensile setting of microsize is used to compare \rrevC{results of} a GP model with DDD results. {\rev The GB modeling in the DDD simulations is the same as in the present work}, i.e., dislocation glide is restricted to remain within \rrevC{the grains}. {\rev Microhard boundary conditions are utilized in the used GP model {\rev of \cite{aifantis2009discrete}}. However, in a reevaluation of the data and GP theory of~\cip{aifantis2009discrete}, \cit{zhang2014} showed that a much better agreement between DDD and GP results is obtained by using GB yielding in the GP model.}\\
The theoretical basis for the GP model of the work at hand is the framework by \cit{wulfinghoff2013gradient} which supplements \cip{wulfinghoff2012equivalent} \rrevC{(see \cip{wulfinghoff2013equivalent} for the time integration algorithm)} by a GB energy and a GB yield condition. \rrtwoC{In the energetic framework used, the GB energy models the storage of energy due to the accumulation of defects at the GBs. The GB yield strength models the resistance of the GBs against plastic flow, necessary to match the strain profiles obtained by DDD in this work.} For numerically efficient computations, the gradient contribution to the free energy is formulated with respect to a scalar equivalent plastic strain~\f{\gaq}, instead of considering gradients of all plastic slips individually. \rrtwoC{Thus, a pragmatic engineering approach rather than a physical approach is taken.} \rrtwoC{Usually, the magnitude of size effects modeled with GP is associated to the internal length scale used in models. \rrtwoC{Such a length scale is introduced in the mentioned framework via a quadratic defect energy (see \cip{wulfinghoff2013gradient}, and \secref{sec:const} for details). Increasing the GB yield strength in this model, however, significantly intensifies the magnitude of size effects observed. Changing the internal length scale has a significantly smaller effect, see \cip{wulfinghoff2013gradient}.}} Additionally, in \cip{wulfinghoff2013gradient}, an isotropic Voce hardening law for the bulk, also posed in terms of~\f{\gaq}, is used. This phenomenologically accounts for hardening associated to mechanisms like dislocation trapping. The combination of these approaches has shown to give good agreement with the size effect results from a tensile test experiment on microwires \cip{wulfinghoff2013gradient}. In addition, it has been shown in \cite{ziemann2015} that the \f{\gaq}-distributions, obtained from simulations of single-crystalline gold microwires, are compatible with the experimentally determined kernel average misorientation distributions in cross sections of these \rrevC{micro}wires.\\
\rtwoC{Due to numerical reasons, a micromorphic approach, suggested by \cit{forest2009micromorphic}, is used for the implementation of the theory at hand. In the micromorphic approach, additional variables are introduced as additional internal degrees of freedom. Micromorphic theories belong to the class of generalized continuum theories. It can be formally shown that gradient theories, which also belong to this class, are special cases of micromorphic theories (see \cip{forest2003elastoviscoplastic}, and also \cip{lazar2010dislocations} for an overview on dislocations in generalized continua frameworks). In \cip{forest2009micromorphic}, a general framework for the micromorphic approach is outlined. It includes balance equations governing the micromorphic degrees of freedom, boundary conditions, and higher-order stresses (see also the references in \cite{forest2009micromorphic}). An overview of applications to, e.g., elasticity and gradient plasticity is also given. In this context, it is described in detail that models formulated with the micromorphic approach are related to existing gradient models. Therefore, the micromorphic variable is constrained to be equal to its (macro) counterpart. This constraint can be imposed by a penalty term in the free energy. The constrained micromorphic approach yields models \rrevC{that}, e.g., belong to the class of gradient of internal variable models \cip{maugin1990internal}. \rroneC{For instance, the gradient theory by \cit{gurtin2003framework} can be regarded as a constrained micromorphic theory \cip{forest2009micromorphic}}.\\
In the work at hand, the GP approach by \cip{wulfinghoff2013gradient}, implemented within a constrained micromorphic approach \cip{forest2009micromorphic}, is extended by GB hardening.} \rrthrC{Fully threedimensional GP simulations are performed.} The GP results are compared to DDD results. This comparison includes GP results obtained by using Voce hardening on one hand, and GP results obtained by using GB hardening on the other hand. Different crystal orientations and, consequently, different slip system mismatches between adjacent grains, are considered.
\topic{Notation} A direct tensor notation is preferred throughout the text. Vectors and 2nd-order tensors are denoted by bold letters, e.\,g., by~\f{\fa} or~\f{\fA}. A linear mapping of 2nd-order tensors by a 4th-order tensor is written as \f{\fA=\ffC[\fB]}. The scalar product and the dyadic product are denoted, e.\,g., by~\f{\fA\cdot\fB} and~\f{\fA\otimes\fB}, respectively. The composition of two 2nd-order tensors is formulated by~\f{\fA\fB}. Matrices are denoted by a hat, e.\,g., by~\f{\hat\varepsilon}.
\section{Gradient Plasticity Model}
\subsection{Motivation for the Grain Boundary Hardening Extension}

% {\rev {\rev A finite GB yield strength is used in conjunction with an additional hardening relation associated to the GBs. This association of hardening to the grain boundary is motivated from observations in DDD simulations where dislocations entangle close to the GBs. As a consequence of the pile-ups, backstresses arise that influence the distribution of plastic strain in the immediate vicinity of GBs.} In order to be able to account for the distribution of plastic strain in the vicinity of the GBs in the GP model the GB hardening relation is introduced. For this purpose the framework of \cite{wulfinghoff2013gradient} is used. {\rev However, the linear grain boundary energy depending on the scalar equivalent plastic strain~\f{\gaq} is supplemented by an additional quadratic term. This leads to a linear hardening contribution in the yield strength of the GB.}}
\roneC{The presence of GBs in microstructured materials leads to dislocation pile-ups that influence the overall work hardening. In many crystal plasticity continuum models, the dislocation-induced hardening mechanisms are modeled by, e.g., isotropic hardening relations for the bulk material. When solely the overall mechanical properties are of interest, this procedure is able to achieve good results. In the work at hand, however, the distribution of plastic strain is evaluated, additionally. It is investigated, if the localization of plastic strain, resulting from pile-ups of dislocations observed at impenetrable GBs in DDD simulations, can be accounted for by a GP model with GB yielding. \rrtwoC{A direct translation of the DDD grain boundary conditions to the GP model is not possible due to the coarsening in the continuum approach.} In order to account for work hardening, at first, a bulk Voce law is used since it has given promising results in combination with GB yielding in a previous work \cip{wulfinghoff2013gradient}. It is shown \rrevC{here}, however, that this approach is not able to capture the evolution of the plastic strain in the vicinity of the GBs. As an alternative \rrevC{to} this bulk hardening model, a GB hardening relation is used, subsequently. This relation takes the equivalent plastic strain at the GBs into account. The explicit association of hardening to the GBs is motivated by observations in the DDD simulations. Dislocations entangle localized as pile-ups close to the GBs. In the context of GB yielding, it is also noteworthy that in \cit{aifantis2006interfaces}, GB yield stresses are estimated based on indentation studies. The authors attributed the observed increase in hardness near the GB (and, thus, an increase in resolved shear stress) to dislocation pile-ups~\cip{soer2005incipient}. Keeping this in mind, the explicit consideration of the GBs in the hardening relation of the GP model is justifiable.}

\subsection{Mathematical Framework for Gradient Plasticity with Grain Boundary Hardening}
\subsubsection{Basic Assumptions}
In a geometrically linear framework for deformations, the strain tensor reads~\f{\feps=\fsym{\nabla\fu}}. The displacement gradient is given by \f{\nabla\fu=\partial_{x_j}u_i\fe_i\otimes\fe_j}, in terms of the basis vectors~\f{\{\fe_1,\fe_2,\fe_3\}} in a Cartesian coordinate system. Furthermore, the plastic strain tensor is
\EE
\feps^{\rm p}=\sum_\alpha\lambda_\alpha\fM_\alpha^s.
\label{epsilon_plast}
\Ee
The symmetric part of the Schmid tensor reads \f{\fM_\alpha^s=\fsym{\fd_\alpha\otimes\fn_\alpha}} with the slip systems~\f{\alpha=1,2,\ldots,N}. Slip directions are indicated by~\f{\fd_\alpha} and the slip plane normals by~\f{\fn_\alpha}. \rtwoC{It should be noted that in the following plastic slip parameters~\f{\lambda_\alpha} are used.} They increase monotonously, i.e., \f{\dot\lambda_\alpha\geq0} in this work. Since the focus is on face-centered cubic (FCC) crystals, the number of slip parameters is~\f{N=24}. Considering 12 slip systems, each with two directionally dependent slips, would be equivalent. Additive decomposition of the strain tensor leads to the elastic strain tensor~\f{\feps^{\rm e}=\feps-\feps^{\rm p}}, \rrevC{where~\f{\feps} is the total strain tensor}. The equivalent plastic strain measure is introduced as (cf. \cip{wulfinghoff2012equivalent})
\EE
 \gaq(\hat\lambda)=\sum_\alpha \int  \dot \lambda_\alpha  \d t = \sum_\alpha\lambda_\alpha.
\label{gammaEq}
\Ee
\subsubsection{Principle of Virtual Power}
The field equations can be derived from the principle of virtual power. It states that the virtual power of the internal forces~\f{\delta \cP_{\rm int}} equals the virtual power of the external forces~\f{\delta \cP_{\rm ext}}. Body forces are {\blue not present} in the following derivations. \rtwoC{The internal power density of the bulk, \f{p_{\rm vol}}, is assumed to be given by}
\EE
% p_{\rm int}=\fsigma\cdot\dot\feps+\varsigma\dgaq+\fxi\cdot\nabla\dgaq.
p_{\rm vol}=\fsigma\cdot\dot\feps+{\rev \pi}\dcaq+\fxi\cdot\nabla\dcaq.
\Ee
\rtwoC{This statement is an extension of the classic power of internal forces. It takes the power expended by generalized stresses, via the rate of a micromorphic field variable~\f{\zeta} and via the rate of its gradient, into account \cip{forest2010some}}. The stresses~\f{\fsigma}, \f{\rev \pi}, and \f{\fxi} are work conjugate to~\f{\dot\feps}, \f{\dcaq}, and~\f{\nabla\dcaq}, respectively, \rrtwoC{where the scalar micromorphic variable~\f{\zeta} is a different quantity than the magnitude~\f{\xi} of the vectorial microstress~\f{\fxi}.} \rtwoC{The body~\f{\cB} has an external boundary~\f{\partial\cB} and internal boundaries \rrevC{denoted by} the union~\f{\Gamma} of all GBs.} \rtwoC{Then, the internal power~\f{\cP_{\rm int}} is assumed to be given in terms of the volume integral over the internal {bulk} power density~\f{p_{{\rm vol}}} and the area integral over the power density on the {union~\f{\Gamma} of all} GBs, i.e.,}
\EE
% \cp
% \cP_{\rm int}=\int\limits_\cB p_{\rm int}\d V+\int\limits_\Gamma\Xi_\Gamma\dgaq\d a,
\cP_{\rm int}=\int\limits_\cB p_{\rtwo{\rm vol}}{\rtwo \d v}+\int\limits_\Gamma\Xi_\Gamma\dcaq\d a.
\Ee
\rrtwoC{The GB microtraction~\f{\Xi_\Gamma} is associated to~\f{\dcaq} on the GBs. In fact, this microtraction is imposed in form of a jump condition for the microstresses on the GBs, as shown below.} \roneC{Furthermore, the micromorphic counterpart~\f{\caq} of the equivalent plastic strain~\f{\gaq} is assumed to be continuous across the GBs, cf.~\cip{wulfinghoff2013gradient}. Consequently, possible jumps in~\f{\caq} are neglected \rrevC{in the following} (see also the discussion in \secref{subsec:assumptions}).}\\
The external power~\f{\cP_{\rm ext}} is assumed to \rrevC{consist of} the following two contributions
\EE
% \cP_{\rm ext}=\int\limits_{\partial\cB}\left(\bar\ft\cdot\dot\fu+\bar\Xi\dgaq\right)\d a,
\cP_{\rm ext}=\int\limits_{\partial\cB}\left(\bar\ft\cdot\dot\fu+\bar\Xi\dcaq\right)\d a.
\Ee
This takes into account the power expended by tractions~\f{\bar\ft} and microtractions~\f{\bar\Xi}, respectively, at the {\rtwo external} boundary~\f{\partial\cB}. \rtwoC{Due to the simplification of considering an equivalent plastic strain~\f{\gaq}, and an additional micromorphic field variable~\f{\zeta} with its gradient~\f{\nabla\zeta}, only one microtraction~\f{\bar\Xi} \rrtwoC{associated} to the rate~\f{\dot\zeta} is \rrevC{accounted for} on the external boundary. \rrtwoC{This microtraction is imposed as a boundary condition for the microstress, as it is shown below.} \rroneC{In models that account for the individual slips or dislocation densities as field variables it is, in principle, possible to prescribe microtractions for the individual slip systems. Whether this is applicable depends on the model and the employed higher-order quantities (e.g., \cip{gurtin2005boundary,gurtin2010finite}).}} In the following, \f{\dot\fu=\delta\dot\fu} and \f{\dcaq=\delta\dcaq}, where~\f{\delta\dot\fu} and~\f{\delta\dcaq} are arbitrary virtual rates that vanish at the Dirichlet boundaries~\f{\partial\cB_{\rm u}} for given~\f{\{\fu,\caq\}}. The virtual power of the external forces is then given by
\EE
\delta\cP_{\rm ext}=\int\limits_{\partial\cB_{\rm t}}\bar\ft\cdot\delta\dot\fu\d a+\int\limits_{\partial\cB_{\Xi}}\bar\Xi\,\delta\dcaq\d a,
\label{eq:delta_p_ext}
\Ee
with~\f{\partial\cB=\partial\cB_{\rm t}\cup\partial\cB_{\Xi}\cup\partial\cB_{\rm u}} and \f{\partial\cB_{\rm t}\cup\partial\cB_{\Xi}\cap\partial\cB_{\rm u}=\O{}}.
For the same choice of virtual rates, the virtual power of the internal forces, consequently, reads
\EE
% \delta\cP_{\rm int}=\int\limits_\cB \left(\fsigma\cdot\delta\dot\feps+{\rev \pi}\delta\dgaq+\fxi\cdot\nabla\delta\dgaq\right){\rev \d v}+\int\limits_\Gamma\Xi_\Gamma\delta\dgaq\d a,
\delta\cP_{\rm int}=\int\limits_\cB \left(\fsigma\cdot\delta\dot\feps+{\rev \pi}\delta\dcaq+\fxi\cdot\nabla\delta\dcaq\right){\rev \d v}+\int\limits_\Gamma\Xi_\Gamma\delta\dcaq\d a.
\label{eq:delta_p_int}
\Ee
\rtwoC{Here, it should be noted that the variables~\f{\feps^{\rm p}} and~\f{\zeta} are, a priori, chosen to be independent, i.e., \f{\delta\dot\feps^{\rm p}=0} (cf.~\cip{wulfinghoff2014numerically}).} \rtwoC{With~\f{\delta\cP_{\rm int}=\delta\cP_{\rm ext}} and a standard procedure (e.g., \cip{gurtin2007gradient}), it is possible to derive the field equations, the Neumann boundary conditions, and the GB conditions (see also \cip{wulfinghoff2012equivalent,wulfinghoff2014numerically}).} \rtwoC{In Box~1, the resulting field equations are summarized. The classic balance of linear momentum is supplemented by a microforce balance. This takes into account the microstresses occurring due to the introduction of additional contributions to the internal power density. These terms consider the rates of the micromorphic field variable~\f{\zeta}, and its gradient~\f{\nabla\zeta}. Consequently, additional microtraction conditions for the gradient stress~\f{\fxi} at the GBs and at the external boundary supplement the Neumann BC of the Cauchy stress.}
% \TT{htbp}
\begin{figure*}[htbp]
\caption*{Box 1: Field equations and boundary conditions. The jump of~\f{\fxi} is denoted by~\f{\llbracket\fxi\rrbracket=\fxi^+-\fxi^-}. The grain boundary normal points from ``-'' to ``+''.}
\centering{
\renewcommand{\arraystretch}{1.2}
\TB{|lrll|}
\hline
% \hline
Linear momentum balance   & \f{\fzero}        & \f{=\div{\fsigma}}&\f{\quad\forall\fx\in\cB}\\
%   Microforce balance        & \f{\check p}&\f{=\beta-\div{\fxi}=-\partial_{\gamma_{\rm eq}} W_\chi;\quad\beta=\partial_{\gamma_{\rm eq}} W_{\rm h}}\\
  Microforce balance& \f{{\rev \pi}}&\f{=\div{\fxi}}&\f{\quad\forall\fx\in\cB\setminus\Gamma}\\
  GB microtraction& \f{\Xi_\Gamma}&\f{=\llbracket\fxi\rrbracket\cdot\fn}& \f{\quad\forall\fx\in\Gamma}\\
  Neumann BCs for: Cauchy stress            & \f{\fsigma \fn}     & \f{= \bar\ft} & \f{\quad\mathrm{on}\ \partial \cB_t}  \\
		\phantom{Neumann BCs for: }Gradient stress &\f{\fxi \cdot \fn}	&\f{= \bar \Xi} &\f{\quad\mathrm{on} \ \partial \cB_{\,\Xi}}
		\\\hline
% 		\hline
\Tb
}
\end{figure*}
\subsubsection{Constitutive Assumptions}\label{sec:const}
\paragraph{{\rev Free} energy density}
Following the model of~\cip{wulfinghoff2013gradient}, the {\rev free} energy density in the bulk is assumed to have an elastic (\f{W_{\rm e}}), an isotropic hardening (\f{W_{\rm h}}), and a defect~{\rone (\f{W_{\rm g}})} contribution, i.e.,
% \EE
% W(\feps,\feps^{\rm p},\gaq,\nabla\gaq)=W_{\rm e}(\feps,\feps^{\rm p})+W_{\rm h}(\gamma_{\rm eq})+W_{\rm g}(\nabla\gamma_{\rm eq}).
% \label{eq:stored_energy}
% \Ee
\EE
{\rone
W(\feps,\hat\lambda,\caq,\nabla\caq)=W_{\rm e}(\feps,\feps^{\rm p}(\hat\lambda))+W_{\rm h}(\caq)+W_{\rm g}(\nabla\caq)+\roneC{W_{\chi}(\zeta-\gamma_ {\rm eq}(\hat\lambda))}.
}
\label{eq:stored_energy}
\Ee
\rrtwoC{The need for additional contributions to the free energy arises from the coarsening in the continuum modeling of the elastic energy. However, distinct dislocation phenomena are modeled here using the two contributions \f{W_{\rm h}(\caq)} and~\f{W_{\rm g}(\nabla\caq)}: on one hand, the isotropic hardening resulting from statistically stored dislocations, and, on the other hand, the influence of geometrically necessary dislocations.}\\
\rtwoC{The coupling of~\f{\zeta} and~\f{\gaq} is achieved \rrevC{by using} \f{W_\chi(\zeta-\gamma_ {\rm eq}(\hat\lambda)) = H_\chi(\zeta-\gamma_{\rm eq})^2/2}. This penalty energy ensures that~\f{\zeta\approx\gaq} for a sufficiently large penalty parameter~\f{H_\chi}. Thus, in~\eqreff{eq:stored_energy}, the gradient extension is also performed in terms of the micromorphic field variable~\f{\zeta}, instead of~\f{\gaq}.}{\ortb\ The dependence of the {\rev free} energy on the accumulated plastic slip represents a rather simple phenomenological approach. This slip does not represent a state variable describing the internal defect structure locally.} The {\rev elastic energy density reads~\f{W_{\rm e}(\feps,\feps^{\rm p}(\hat\lambda))=(\feps-\feps^{\rm p}(\hat\lambda))\cdot\ffC[\feps-\feps^{\rm p}(\hat\lambda)]/2}.}
% \begin{eqnarray*}
%  W_{\rm e}(\feps,\feps^{\rm p})&=&1/2(\feps-\feps^{\rm p})\cdot\ffC[\feps-\feps^{\rm p}],\\
% W_{\rm h}(\gaq)&=&(\tau_\infty^{\rm C}-\tau_0^{\rm C})\gaq+1/\Theta(\tau_\infty^{\rm C}-\tau_0^{\rm C})^2\exp\left(-\Theta\gaq/(\tau_\infty^{\rm C}-\tau_0^{\rm C})\right),\\
% W_{\rm g}(\nabla\gaq)&=&1/2K_{\rm G}\nabla\gaq\cdot\nabla\gaq.
% \end{eqnarray*}
% \EE
% \rev
%  
% \Ee
Here, \f{\ffC} denotes the elastic stiffness tensor. \rtwoC{Additionally, an isotropic, rate-independent, Voce hardening relation is used via
\EE
W_{\rm h}(\caq)=(\tau_\infty^{\rm C}-\tau_0^{\rm C})\caq+\frac{1}{\Theta}(\tau_\infty^{\rm C}-\tau_0^{\rm C})^2\exp\left(-\frac{\Theta\caq}{(\tau_\infty^{\rm C}-\tau_0^{\rm C})}\right),
\Ee
with the initial yield stress~\f{\tau_0^{\rm C}}, the saturation stress~\f{\tau_\infty^{\rm C}}, and {the initial} hardening modulus~\f{\Theta}.} \roneC{This hardening energy is obtained by an integration of a chosen relation for the hardening stress of all slip systems. Instead, it is possible to include a dissipative hardening contribution on each slip system. This \rrevC{approach} would, however, lead to the same material behavior.} \rtwoC{The defect energy reads}
\EE
W_{\rm g}(\nabla\caq)=\frac{1}{2}K_{\rm G}\nabla\caq\cdot\nabla\caq,
\label{eq:defect_energy}
\Ee
\rtwoC{and introduces a length scale into the model by means of the defect parameter~\f{K_{\rm G}} (which is assumed to be a constant, here). The quadratic formulation of the defect energy gives a linear dependence of the microstress~\f{\fxi} on the gradient~\f{\nabla\zeta}. This gradient could be interpreted as an approximative measure for geometrically necessary dislocation (GND) densities \cip{wulfinghoff2013gradient}. \rroneC{This simplification of representing dislocations using an equivalent measure is only possible for monotonic loading processes since \f{\dot\caq\geq0}.} By the quadratic choice in \eqreff{eq:defect_energy}, the well-posedness of the boundary problem, with respect to the defect energy, is ensured. In this context, see the discussion of a quadratic defect energy function in \cip{reddy2011role}. There, however, the discussed defect energy depends on the sum over the slip gradients rather than on the gradient of the micromorphic counterpart of the sum over the slip parameters.} \roneC{Additionally, the gradient stresses resulting from \eqreff{eq:defect_energy} could also be introduced within a dissipative framework rather than in an energetic one (see, e.g., \cip{fleck2009mathematical} for a fundamental discussion on the matter of dissipative and energetic frameworks).}\\
In addition to the bulk {\rev free} energy density~\f{W}, an energy density per unit surface has been introduced on the GBs in \cip{wulfinghoff2013gradient}
\EE
\label{eq:gb_energy}
W_\Gamma(\caq)=\Xi_0^{\rm C}\caq.
\Ee
{\rone In combination with a GB yield criterion, this leads to an explicit consideration of GB yielding in the continuum model, cf.\ also the brief discussion of GB energies in the introduction.% \rrevC{of the work at hand}.}
% , where it has been shown in the above mentioned work that the resulting finite GB yield strength from this approach is then the dominating influence on the size effects observed with this model.
\paragraph{Hardening extension of energy density on the grain boundary}
{\rtwoC{The GP model is extended to account for hardening considering the plastic deformation of the GBs. Therefore, the following extension of the GB energy, \eqreff{eq:gb_energy}, is proposed}}
\EE
W_\Gamma(\caq)=\Xi_0^{\rm C}\caq+\frac{1}{2}K_{\rm H}\caq^2,
\label{eq:gb_energy_hard}
\Ee
\rtwoC{where~\f{K_{\rm H}} is a GB hardening parameter. Contrary to \eqreff{eq:gb_energy}, this GB energy leads to an increasing GB yield strength in the additional GB yield criterion used in this work. Consequently, the GB yield strength increases in dependence of~\f{\caq}. This mechanism phenomenologically accounts in the continuum model for the strengthening effects caused by the pile-ups of dislocations at the impenetrable GBs in the discrete model.}
\paragraph{Dissipation}
Upon neglecting thermal effects, the total dissipation reads
\EE
D_{\rm tot} = \cP_{\rm ext} - \int \limits_\cB \dot W \d v - \int \limits_{\Gamma} \dot W_\Gamma \d a \geq 0.
\label{eq:dissipation_total}
\Ee
After exploiting \f{\cP_{\rm ext}=\cP_{\rm int}}, it can be summarized as
\EE
\label{eq:dissipation_total_D}
 D_{\rm tot} = \int \limits_\cB \cD \d v + \int \limits_{\Gamma} \cD_\Gamma \d a \geq 0.
\Ee
Substitution of~\eqreff{eq:stored_energy} in~\eqreff{eq:dissipation_total} gives a bulk dissipation {\rev in~\eqreff{eq:dissipation_total_D} that reads}
\EE
\label{eq:bulk_dissipation}
%  \cD = (\fsigma - \partial_{\blds{\feps}} W_{\rm e}) \cdot \dot \feps - \partial_{\blds \feps^{\rm p}} W_{\rm e}  \cdot \dot \feps^{\rm p}+ ({\rev \pi} - \underbrace{\partial_{\gaq} W_{\rm h}}_\beta) \dgaq + (\fxi - \partial_{\, \nabla \gaq} W_{\rm g}) \cdot \nabla \dgaq \geq0.
 \cD = (\fsigma - \partial_{\blds{\feps}} W_{\rm e}) \cdot \dot \feps - \partial_{\blds \feps^{\rm p}} W_{\rm e}  \cdot \dot \feps^{\rm p}+ ({\rev \pi} - {\partial_{\caq} W_{\rm h}}{\rone-\partial_{\zeta}W_\chi}) \dcaq {\rone -\partial_{\gaq} W_\chi\dgaq} + (\fxi - \partial_{\, \nabla \caq} W_{\rm g}) \cdot \nabla \dcaq \geq0.
\Ee
{\rone
Introducing the abbreviation~\f{\check p=\partial_{\gaq} W_\chi=- \partial_\zeta{W_\chi}}, assuming the stresses~\f{\fsigma},~\f{\pi}, and~\f{\fxi} to be purely energetic, and substituting~\f{\partial_{\feps^{\rm p}}{W_{\rm e}}=-\fsigma} leads to a reduced bulk dissipation inequality of the form}
\EE{\rone
 \cD = \fsigma \cdot \dot \feps^{\rm p} - \check p\, \dot \gamma_{\rm eq}  \geq 0.}
\label{eq:diss_reduced}
\Ee
{\rtwo Resulting from the micromorphic approach, the (reduced) dissipation inequality includes a stress~\f{\check p} that is associated to the difference \f{\gaq-\zeta} (see also \cip{forest2009micromorphic}).}
% In \cip{forest2009micromorphic} such a quantity is termed ``the thermodynamic force associated with the internal variable``.}
It is assumed that the bulk dissipation is induced by the dissipative shear stresses~\f{\tau_\alpha^{\rm d}} of the individual slip systems, e.g., \cip{cermelli02},
\EE
\cD=\sum_\alpha\tau_\alpha^{\rm d}\dot\lambda_\alpha.
\label{eq:dissipation_induced}
\Ee
Consequently, after combining~\equreff{epsilon_plast}{gammaEq} with~{\rone\equreff{eq:diss_reduced}{eq:dissipation_induced}}, using the microforce balance equation from Box~1 as well as the abbreviation~\f{\beta = \partial_{\zeta}W_{\rm h}}, the dissipative shear stresses read
\EE
\tau_\alpha^{\rm d}=\tau_\alpha+\div\fxi-\beta,
\Ee
with the resolved shear stresses~\f{\tau_\alpha = \fsigma \cdot \fM^{\rm S}_\alpha}, and the hardening microstress~\f{\beta}.\\
Furthermore, the GB dissipation {\rone from~\eqreff{eq:dissipation_total} and~\eqreff{eq:dissipation_total_D}, respectively,} reads
\EE
 \cD_\Gamma = (\Xi_\Gamma - \Xi_\Gamma^{\rm e})\, \dcaq = \Xi_\Gamma^{\rm d}\, \dcaq \geq 0,
\Ee
where~\f{\Xi_\Gamma^{\rm d}} is the dissipative GB microtraction. The energetic GB microtraction~\f{\Xi_\Gamma^{\rm e}} reads
\EE
\Xi_\Gamma^{\rm e}= \partial_{\caq} W_\Gamma=\Xi_0^{\rm C}+K_{\rm H}\caq.
\label{eq:gb_microtraction}
\Ee
\paragraph{Flow rules for the bulk and the grain boundary}
The flow rule for the bulk is assumed to be of overstress type, {\rev formulated in the slip parameter} rates~\f{\dot\lambda_\alpha},
\EE
\label{eq:powerLaw}
 \dot \lambda_\alpha = \dot \gamma_0 \left\langle
\frac{\tau_\alpha^{\rm d} - \tau^{\rm C}_0}{\tau^{\rm D}}
\right\rangle^p
=\dot \gamma_0 \left\langle
\frac{\tau_\alpha+\div\fxi- ( \tau^{\rm C}_0 + \beta )}{\tau^{\rm D}}
\right\rangle^p,
\Ee
with the reference shear rate~\f{\dot\gamma_0}, the drag stress~\f{\tau^{\rm D}}, and the rate sensitivity exponent~\f{p}. The GB yield function is introduced as
\EE
 f_\Gamma = \Xi_\Gamma^{\rm d} = [\![ \fxi ]\!] \cdot \fn -\Xi_\Gamma^{\rm e}{\redsw = [\![\fxi]\!]\cdot \fn - (\Xi_0^{\rm C}+K_{\rm H}\caq)},
\label{eq:gb_yield_function}
\Ee
where the GB itself is assumed to deform dissipation-free. However, in principle, it is possible to additionally consider GB dissipation.
% The energetic GB microtraction in~\eqreff{eq:gb_yield_function} is given by~\eqreff{eq:gb_microtraction}.
Contrary to a GB yield strength resulting from \eqreff{eq:gb_energy}, the GB yield strength {\rone stemming} from \eqreff{eq:gb_energy_hard} combines a constant initial yield strength with an additional contribution linearly increasing with~\f{\caq}, i.e., with proceeding plastic deformation. For the GB, the Kuhn-Tucker conditions read
% \EE
\f{{f_\Gamma\leq0,\,\dcaq\geq0,\,\dcaq f_\Gamma=0}},
% \label{eq:kuhn_tucker}
% \Ee
where the GB is assumed to behave rate-independent.
% {\redsw \subsection{Finite Element Implementation}
% It is noted that the GP theory at hand is implemented with finite elements in an in-house code. For details on the implementation, also considering GB energies of orders higher than one it is referred to~\cit{wulfinghoff2013gradient}. The enhanced time integration algorithm is described in~\cit{wulfinghoff2013equivalent}.}
\section{Discrete Dislocation Dynamics Simulations}
\label{sec:ddd}
\subsection{Introductory Notes}
As a data basis for the parameter calibration of the GP model, a multitude of DDD simulations is carried out. The discrete dislocation results are obtained with the DDD code described in detail in \cip{Siska2009,Weygand2002,Weygand2009}, \rrevC{including the analysis of the elastic interactions of dislocations as well as the resulting plastic deformation of a sample.} {\rev The framework allows simulating polycrystalline aggregates.}
% In the following the used DDD model is briefly described and relevant works from the literature are referenced. 
{\rev For brevity, only a short description of the framework is given. Material parameters, interface and boundary conditions, and the simulation setup are outlined. The necessary ensemble averaging procedure is described by which simulation results directly comparable to the (non-scattering) GP simulation results are obtained.}
% i
\subsection{Material Model}
\label{subsec:ddd}
{\roneC{In the DDD code, an individual crystallographic orientation is assigned to each grain and isotropic linear elasticity is used.}} An FCC crystal system with elastic constants mimicking aluminum (shear modulus {\rthr \f{G=27}~GPa}, Poisson's ratio {\blueeb $\nu=0.347$)} is used. \roneC{The initial dislocation configurations are chosen such that the system is placed in the multiplication controlled regime \cip{Kraft2010}. Frank-Read (FR) sources are, therefore, distributed randomly with respect to position and orientation. \rrevC{This procedure is performed under the restriction that each slip system in each grain contains} the same number of sources. The initial source length is chosen to vary between $0.16-0.27$~\microm\ in order to reduce artificiality that would be introduced into the model by the choice of a uniform FR source length. Using this procedure, the initial dislocation density is about $\rho\approx7.5\times10^{13}\,$m$^{-2}$. \rrevC{The sources are approximately two times larger than the mean dislocation spacing~\f{\ell=1/\sqrt{\rho}\approx0.12}~\microm,} \rrevC{which places the system behavior in the} multiplication controlled plasticity regime \cite{Kraft2010,zhang2014}. Consequently, the work hardening is influenced by dislocation reactions, as opposed to single source controlled plasticity. Thus, a comparison of DDD results with results from a continuum model is feasible.}
\subsection{Simulation Setup, Geometry and Boundary Conditions}
\label{subsec:ddd_sim_setup}
Strain rate controlled tensile tests are simulated along the $\langle 100 \rangle$-axis (which is the $x$-axis of the in-lab frame) {\rev of a tricrystal geometry} (cf.~\figref{fig:gp_vs_ddd_fem}) with a strain rate of $\dot{\varepsilon}=5000\,$s$^{-1}$ {\orange during the simulation time~$t$.} This \rrevC{rate is chosen} {\rev due to the small time scale at which DDD operates. It is assumed that the plastic behaviour is independent of the strain rate~\cite{Senger2008}.} {\blue In order to investigate different dislocation interaction behavior across the GBs in the model, the crystal orientation of the central grain is rotated by an angle~\f{\varphi} around the $x$-axis (cf.\ also~\secref{sec:voce_nlc}).} {\rev The size of the cubic grains} is set to $0.75$~\microm\ and {\rev the BCs applied are as follows.}
% \begin{itemize}
%  \item[(i)] $u_x(x=0)=0$ and prescribed $u_x(x=x_{\rm max})=x_{\rm max}\dot{\varepsilon}t$ while the {\blue$y$- and $z$-directions}, orthogonal to the loading axis, are traction free on these planes (with $x={\rm const.}$). {\blue {\rev For all} non-$x$-components of {\rev dislocation Burgers vectors}, the boundaries are transparent, while for the $x$-components, the finite element part of the DDD code exerts an element size (and therefore discretization-) dependent backstress. In the limiting case of an infinitesimal finite element mesh, the boundaries would essentially behave like an impenetrable grain boundary. {\rev In the following this Burgers vector dependent transparency condition is called semi-transparent BC.}} Furthermore, translation in the $y$-$z$-plane and rotation around the $x$-axis are suppressed.
{\rev On the two bounding planes (in loading direction) the displacements \f{u_x(x=0)=0} and \roneC{\f{u_x(x=x_{\rm max})=x_{\rm max}\dot{\varepsilon}t}}, respectively, are prescribed} while the displacements orthogonal to the loading axis are set to zero {\rev (except for one special case, cf.\ \secref{subsec:LC0E})}. 
% \end{itemize}
All other boundaries are traction free and dislocations are allowed to leave the volume, there.
% {\rev The DDD simulations {\rev naturally} yield scattering results, e.g., in the stress-strain curves. Thus, a suitable averaging procedure has to be used in order to allow one to calibrate GP model parameters to these results.}
\subsection{Averaging Procedure}
\label{sec:ddd_averaging}
{\rev % % % % % % \subsection{Averaging Procedure due to Scattering of Discrete Simulations}
Due to the discrete nature of DDD, the scatter in the results depends on the dislocation structure and density. The higher the density, the closer DDD results from individual simulations come to a continuum-like profile. However, the size of the setups considered in the present work is within the size effect regime, where individual dislocations control plasticity. Thus, a suitable averaging procedure is necessary as a prerequisite to calibrate GP model parameters to these results. Therefore, averaging over an ensemble of simulations is performed.}
Once all DDD simulations are carried out, the plastic strain of each realization is directly evaluated {\blue from the swept areas of all discrete dislocations}. This postprocessing procedure considers slices perpendicular to the tensile axis. {\oreb Subsequently, the DDD results from a number of simulations~$M$ are averaged, for one set of BCs and crystal orientations:
% . The averaging procedure is described in the following.
\begin{enumerate}
 \item The contribution of a slip system~\f{\alpha} to the plastic strain tensor is calculated in each slice {\rev of \rrevC{the} volume $V$} via $\feps^{\rm p}_\alpha = bA_\alpha/(2V)\left(\fd_\alpha\otimes\fn_\alpha + \fn_\alpha\otimes\fd_\alpha\right)$, where $b$ is the length of the Burgers vector, {\rev and~\f{A_\alpha} the swept area of dislocations of slip system~\f{\alpha}}.
 \item The slip system contributions~\f{\feps^{\rm p}_\alpha} to the plastic strain tensor~\f{\feps^{\rm p}} of a slice are superposed.
 \item The mean plastic strain per slice is obtained by averaging twice:
 \begin{enumerate}
  \item Averaging the plastic strain arithmetically, i.e., over $\sim 5$ DDD simulations.
  \item \rrtwoC{Averaging the plastic strain over equivalent volumes regarding the crystallographic mirror-symmetry in $x$-direction at \f{x=0.5\,x_{\rm max}}. Thus, the data from DDD is doubled and, therefore, further smoothed. This is, formally, similar to the averaging over all symmetry-equivalent components of the dislocation density tensor and symmetry-equivalent positions along the $x$-axis in \cite{aifantis2009discrete}.}
% , i.e., {\blue the mirror-symmetry at \f{x=0.5\,x_{\rm max}} is used to average along the $x$-axis.}} \rrtwoC{The second averaging of the plastic strain is done over equivalent volumes due to the crystallographic symmetry: in $x$-direction, the sample possesses a mirror-symmetry. By using this procedure, the available data from DDD is doubled and therefore further smoothed.}
 \end{enumerate}
\end{enumerate}
}
{\oreb The number of slices along the tensile axis is chosen to be~$150$, resulting in a slice thickness of {\rev $15\,$nm}. This is based on the evaluation of the mean dislocation spacing in the pile-up close to the GBs of approximately $1/\sqrt{\rho}\approx 25\,$nm. With~$150$ slices, the evaluation resolution is {\rev almost} two times higher than the mean dislocation spacing {\rev in the pile-up}, abundant to capture the strain gradients. The GB itself does not exhibit plastic strain, but the slices adjacent to it do. The averaging procedure renders the spatial distribution of plastic strain to be one-dimensional. All dislocations add to the produced plastic slip -~and thus to the plastic strain~- regardless of their locations in the cross sections. The averaged DDD results are used to calibrate the GP model parameters.
}

% . However, the size of the setups considered in the present work is within the size effect regime, where individual dislocations control plasticity. Therefore, each strain profile has to be averaged over $\sim 5$~DDD simulations as well as crystal symmetry, leading to a total number of $10$~DDD simulations. {\rev These serve} as a basis to come to a profile comparable with results obtained from a continuum theory. The number of cross sections, and therefore the resolution of postprocessing, chosen for the evaluation of the strain profiles, is based on the mean dislocation spacing in the pile-ups at the GBs (cf.~\secref{sec:ddd_averaging}). From the choice of cross section number in principle, one might expect a discretization dependence of the plastic strain obtained for the immediate vicinity of GBs. However, it can be shown by refining this discretization by using finer slices that the plastic strain value close to the GB remains nearly the same in the DDD simulations, if the discretization resolution is chosen higher than the dislocation spacing in the pile-ups.\\
% Additionally, the averaging of plastic slip over cross sections along the sample leads to an averaged quantity, because the locations of the dislocations within the cross section become irrelevant: Each dislocation regardless of its specific location in the cross section of a slice adds to the produced plastic slip -~and thus to the plastic strain~- which in turn becomes a one dimensional measure along the tensile direction of the work at hand.}
\section{Gradient Plasticity Simulation Setup}
\subsection{Geometry, Boundary Conditions and Crystal Orientations}
\label{subsubsec:geometry}
{\rev In this section, the setup of the finite element (FE) simulations for the GP model with an in-house code is discussed (for details on the FE implementation see \cite{wulfinghoff2013gradient,wulfinghoff2013equivalent}).}
In all simulations, a tricrystal composed of $0.75$~\microm\rrevC{-wide} cubic grains is considered under tensile loading with Dirichlet conditions (\f{\Delta u_x=0.005L_0}, \f{L_0=2.25}~\microm, cf.~\figref{fig:gp_vs_ddd_fem}). {\rev Lateral contraction on the boundary planes at \f{x=0}, and at~\f{x=x_{\rm max}}, respectively, is prohibited (except for one special case, cf.\ {\rev \secref{subsec:LC0E}}). The restriction of lateral contraction (see also~\figref{fig:gp_vs_ddd_fem}) is {\rev abbreviated by NLC in the following.} At the beginning of the simulation, the equivalent plastic strain~\f{\gaq} {\rev (and its micromorphic counterpart~\f{\zeta})} are set to zero everywhere. Finite element nodes on the GB planes as well as on the boundary planes are {\rev assigned a GB yield strength} and are set to microhard behavior, at the beginning. Once the (GB) yield condition at these nodes is fulfilled, they are allowed to yield.{\rev\ The motivation for assigning a GB yield strength to the boundary planes, too,} is drawn from the BCs in the DDD simulations {\rev that lead to pile-ups of dislocations at the boundaries, as well (cf.~\secref{subsec:ddd})}.}{\rev\ For details on the active set search for the (grain) boundary nodes in the GP simulations it is referred to \cip{wulfinghoff2013gradient}.}\\
\FF{htbp}
\includegraphics[width=\linewidth]{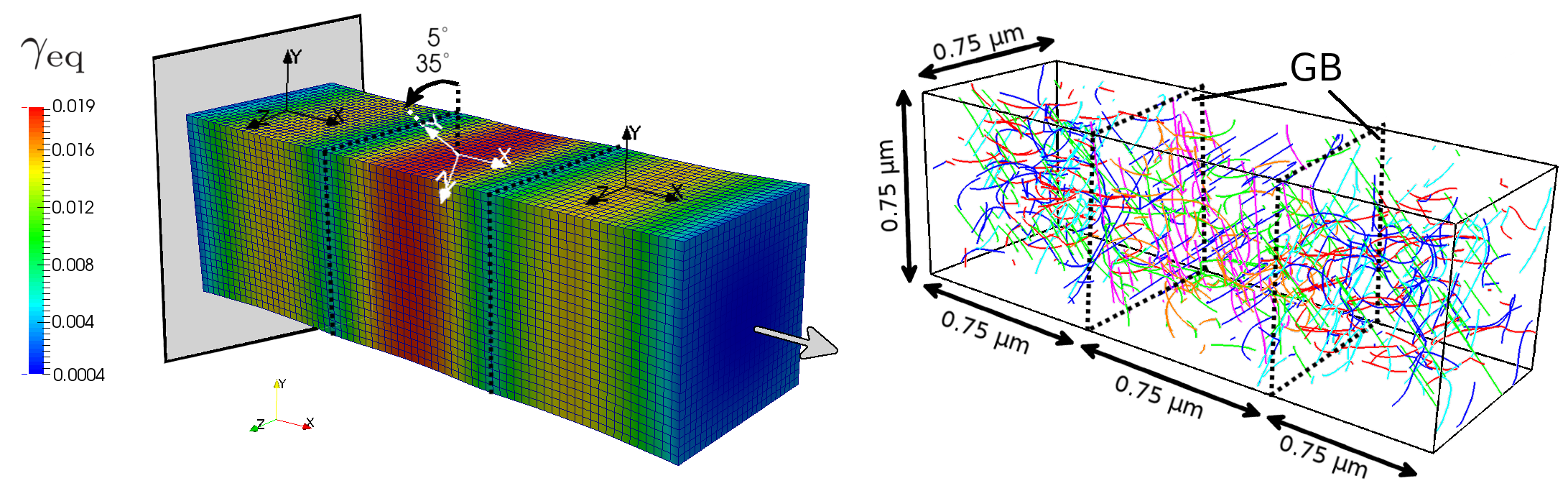}
\FC{Equivalent plastic strain field obtained in a gradient plasticity FEM simulation without lateral contraction at the boundaries and finite values of the GB yield strength (left, depicted are the Gauss point subvolumes). DDD simulation with impenetrable GBs (right, dashed lines indicate the GBs)}
\label{fig:gp_vs_ddd_fem}
\Ff
{\blue An elastically isotropic, {\rev but plastically anisotropic} aluminum-like material is considered throughout {\rev the following} GP simulations. \rrevC{At first, the $\langle 100 \rangle$-orientation of the three grains with FCC crystal structure is oriented aligned with the $x$-axis of the cartesian \f{x,y,z}-system.} The central grain is rotated by an angle~\f{\varphi}, subsequently (see also \secref{subsec:ddd_sim_setup}). As a first approach to investigate the interaction behavior across GBs with both models, three representative cases, \f{\varphi\in\{0\degree,5\degree,35\degree\}}, are selected. Thus, the ideal case of vanishing mismatch between the slip systems of adjacent grains (and, consequently, unrestricted interaction of dislocations across the GBs) is supplemented with a case of small mismatch (yet still strong interaction) and a case of large mismatch (weak interaction) between adjacent grains.}\\
The chosen finite element mesh for the GP simulations consists of \f{12\times12\times12} elements for each grain, i.e., in total \f{25012}~DOF. It is chosen as a compromise between computational time and accuracy. \rtwoC{Compared to the chosen discretization, a refinement of the mesh with approximately twice the number of DOF yields a relative error \rrevC{in} the stress-strain response at the final time-step of less than~\f{0.01}.}
\subsection{Model Parameters}\label{sec:parameters}
{\rev The GP model parameters are calibrated} such that the overall mechanical response matches the averaged stress-strain results of the DDD simulations. This consideration is not sufficient alone since distributions of the plastic strain, e.g., along a line segment of the whole volume, are not necessarily predicted correctly. Therefore, the local distribution of plastic strain is taken into account in the calibration, too. While, in principle, the determination of the GP model parameters in the fitting procedure is not unique, there are some guidelines outlined in this work that should help in their calibration. They are discussed in the following {\rev with \rrevC{regard to} the DDD results as data source. For each case (cf.~\tabref{tab:setups})}
\EN{1.}
\ii A least-squares fit (LSF) of the DDD stress-strain curves is obtained.
\ii The cross-section averaged plastic strain profiles along the loading axis are obtained. {\rev This is performed in order to ensure comparability with the (averaged) plastic strain profiles from DDD. The occurring differences within the cross-section distributions of GP results are {\rev relatively} small {\rev and not as pronounced as they are in the DDD simulations}.}
\ii The Young's modulus is calibrated in the GP model in order to match the elastic stiffness obtained from {\rev the LSF}.
\ii The initial yield stress~\f{\tau_{0}^{\rm C}} of the GP bulk model and the initial yield strength~\f{\Xi_0^{\rm C}} of the GBs as well as the initial yield strength~\f{\Xi_{0,\partial\cB}^{\rm C}} of the boundary (planes at~\f{x=0} and~\f{x=x_{\rm max}}) are calibrated using the DDD-LSF and the averaged plastic strain profiles. Therefore, the plastic strain profiles along the loading axis {\rev of GP results and DDD results} are compared. On that account, plastic strain profiles are obtained at three, {\blue representatively} chosen, fixed overall plastic strain values in the well established plastic regime are used, i.e., \f{\varepsilon_{\rm p}\in \left\{0.001,0.002,0.003\right\}}.
\EN{a)}
\ii In case of Voce hardening: The initial hardening modulus~\f{\Theta} and the saturation stress~\f{\tau_{0}^{\infty}} are adjusted to the work hardening behavior of the DDD-LSF.
\ii In case of GB hardening: The GB hardening parameter~\f{K_{\rm H}} is adjusted to give good agreement with the work hardening behavior of the DDD-LSF. Additionally, the evolution of the plastic strain at the GBs is taken into account{\blue, i.e., the GP plastic strain profiles are compared to the (averaged) DDD plastic strain profiles at all three overall plastic strain values \f{\varepsilon_{\rm p}}}.
\En
\En
The calibration yields a uniform Young's modulus of \f{E=65}~GPa for all GP simulations (except for one special case, cf.~\secref{subsec:LC0E}). The slightly lower value of the Young's modulus, compared to the value of~\f{72.7}~GPa used in the DDD model, is due to the anelastic behavior (bow-out of dislocations) from the very beginning of the loading. The Poisson's ratio is kept identical to the DDD simulations (\f{\nu=0.347}).\\
It is remarked that the defect energy parameter~\f{K_{\rm G}}, although in principle introducing an internal length scale into the model, mainly controls the elastic-plastic transition behavior if the GP model with GB yielding is used. Consequently, \f{K_{\rm G}=84\times10^{-6}}~N is chosen such that the GP simulations show similar stress-strain results as the DDD simulations in this regime. \revC{Hardening resulting from the defect energy is negligible compared to the additional hardening relations investigated in the work at hand (cf.~\figref{fig:k_h_study}).}
{\rtwo For all simulations, a reference shear rate of~\f{\dot\gamma_0=10^{-3}/}s, a rate sensitivity exponent of~\f{p=20}, and a drag stress of~\f{\tau^{\rm D}=1}~MPa are used. The used penalty parameter is~\f{H_\chi=10^8}~MPa. All cases considered are summarized in~\tabref{tab:setups} \rrevC{(see~\secref{sec:voce_nlc} for remaining Voce parameters)}.}
\TT{htbp}
\centering
\TC{Setups and model parameters of GP simulations for comparison to DDD results. {\rev The abbreviations NLC and LC indicate if lateral contraction is prevented or allowed for on the boundary planes at~\f{x=0} and~\f{x=x_{\rm max}}. In the special case LC0E, the two bounding grains are purely elastic}. \revC{The gradient hardening contribution is negligible in the investigated NLC cases, see~\figref{fig:k_h_study}}}
\TB{||c|c|c|c|c|c|c||}
\hline
\vphantom{\Big|}Name&Angle \f{\varphi}&Hardening&\f{\Xi_{0,\rm{GB}}^{\rm C}}&\f{\Xi_{0,\partial\cB}^{\rm C}}&\f{K_{\rm H}}&\f{\tau_0^{\rm C}}\\
&&&[N/m]&[N/m]&10$^3$ [N/m]&[MPa]\\\hline
NLC35V&35$\degree$&Voce&3.5&25&-&30.0\\
NLC35G&35$\degree$&GB&1.5&25&1.8&33.5\\
NLC5G&5$\degree$&GB&1.5&25&1.8&33.5\\
LC0E&0$\degree$&Gradient&-&-&-&44.0\\\hline
\Tb
\label{tab:setups}
\Tt
\section{Comparison of Gradient Plasticity Results to Discrete Dislocation Dynamics Results}
\subsection{Gradient Plasticity with Voce Hardening}
\label{sec:voce_nlc}
At first, the framework of~\cite{wulfinghoff2013gradient} is used (without GB hardening) taking into account the {\rev calibration} guidelines described above. Consequently, the averaged plastic strain profiles are evaluated at three constant overall plastic strains (cf.~\figref{fig:nlc35_gbhard}, left) for the GP simulations. They are compared to the averaged plastic strain profiles (\figref{fig:nlc35_gbhard}, right) obtained from the DDD simulations (indicated schematically by blue error bars in \figref{fig:nlc35_gbhard}, left).  {\rev The model parameters for the case at hand (NLC35V) can be found in \tabref{tab:setups}. In addition, the saturation stress reads \f{\tau_\infty^{\rm C}=108.51}~MPa, and the initial hardening modulus is \f{\Theta=1075}~MPa.}
% \begin{center}
% \FF{htbp}\includegraphics[width=\linewidth]{fig_02_ddd_gp_gbhard_nlc35_working_draft_2_V5}
\FF{htbp}\includegraphics[width=\linewidth]{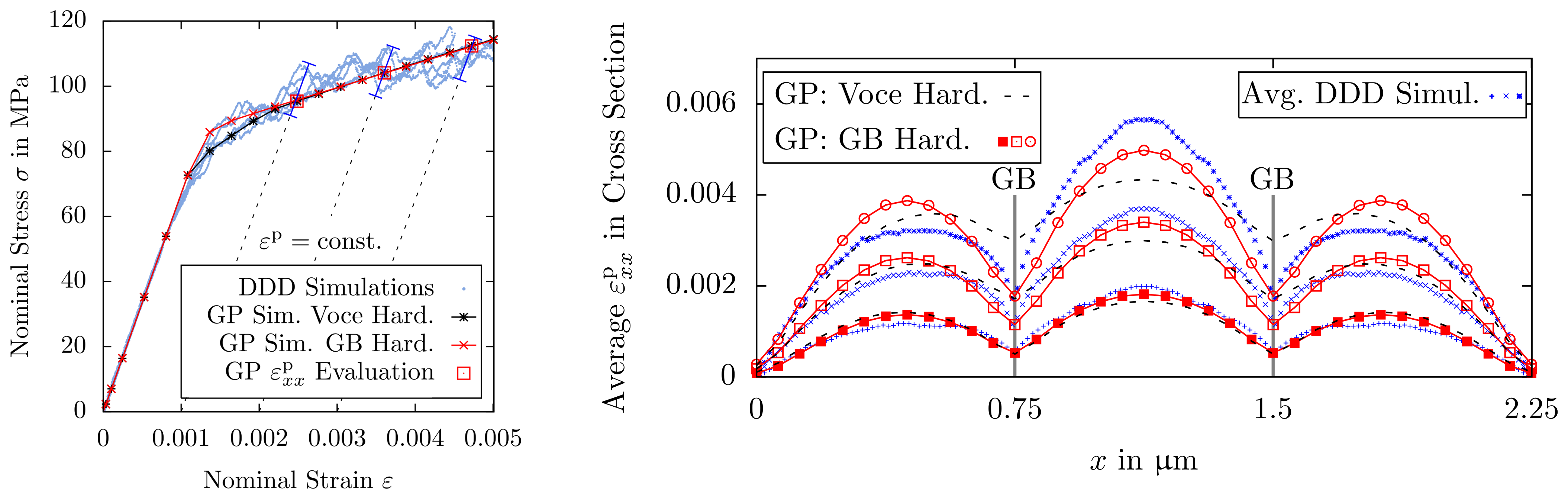}
\caption{NLC35V/NLC35G: Stress-strain curves of DDD and GP simulations (left). Distribution of cross-section averaged plastic strain along loading direction of DDD and GP simulations (right), obtained at fixed overall plastic strain values (left). Boundary conditions with restricted lateral contraction, central grain rotated 35$\degree$ around loading axis.}
%  GB hardening used in the red colored GP results. Voce hardening used in the black colored GP results.}
\label{fig:nlc35_gbhard}
\Ff
It can be seen in~\figref{fig:nlc35_gbhard} that, although the plastic strain profiles for~\f{\varepsilon_{\rm p}=0.001} are in good agreement as a consequence of the calibration of the GB yield strength, the subsequent evolution can not be accounted for by the GP simulations {\blue with Voce hardening}. At the GBs, significant deviations occur which are caused by an obvious limitation of this approach to account for the observed accumulation of dislocations at the GBs in the discrete simulations.
\subsection{Gradient Model with Grain Boundary Hardening}
Voce hardening is not considered in the following, but hardening related to the GBs is (cf.\ NLC35G/NLC5G in \tabref{tab:setups}). It can be seen in~\figref{fig:nlc35_gbhard} and~\figref{fig:nlc5_gbhard} that, {\blue depending on the chosen rotation angle~\f{\varphi} of the central grain,} the strain profiles of the GP simulations are {\rev matching the DDD profiles better,} {\blue either in the central grain, or in the {\rev bounding} grains.}
\FF{htbp}
\includegraphics[width=\linewidth]{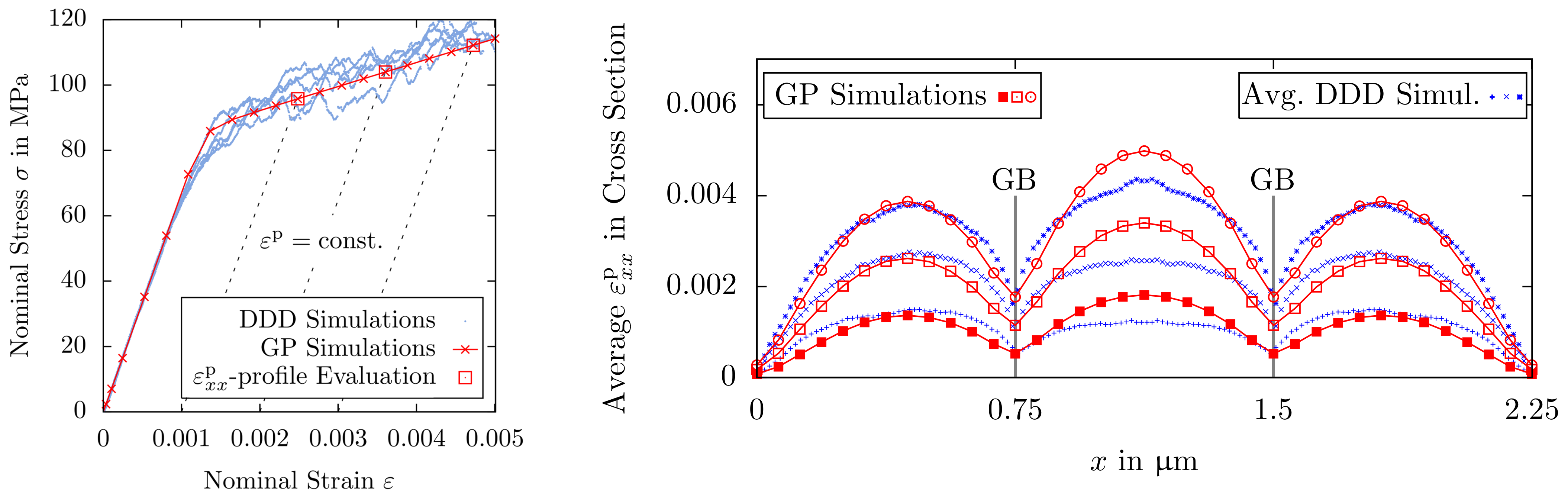}
\caption{NLC5G: Stress-strain curves of DDD and GP simulations (left). Distribution of cross-section averaged plastic strain along loading direction of DDD and GP simulations (right), obtained at fixed overall plastic strain values (left). Boundary conditions with restricted lateral contraction, central grain rotated 5$\degree$ around loading axis. GB hardening used in GP model, no Voce hardening used.}
\label{fig:nlc5_gbhard}
\Ff
{\blue Nevertheless, {\rev in the vicinity of} the GBs, the plastic strain evolution is captured {\rev significantly better compared to the Voce hardening approach.} For both simulations with GB hardening, identical parameters are used (see \tabref{tab:setups}).
\subsection{Gradient Model with Gradient Hardening and without Grain Boundary / Voce Hardening}
\label{subsec:LC0E}
\roneC{It is recalled that in the GP model, gradients enter the theoretical framework via the defect energy, see \eqreff{eq:defect_energy}. In order to isolate the influence of the defect energy on the {evolution of the} strain gradients from the misorientation and the interaction behavior of dislocations across GBs, an additional case is investigated in the following.} Therefore, the two bounding grains are set to be purely elastic. Only the central grain is elastic-plastic and has an ideal $\langle 100 \rangle$-orientation with respect to the $x$-axis (cf.\ \tabref{tab:setups}). {\rev All GB contributions and the Voce hardening contribution are neglected. {\rev In this case, lateral contraction of the bounding planes is not restricted (see \cip{wulfinghoff2013gradient} for the respective BCs).} As a consequence of the isolation of strain gradients, information about the quality of the used defect energy approach in the GP model is obtained}. {\blue Compared to the other investigated settings, the elastic {\rev overall} response of the DDD simulations is harder. This is due the confinement of plastic activity to the central grain. Consequently, for this special case, the Young's modulus used in the GP model has to be slightly adjusted {\rev to a value of 69.4~GPa.} In all other cases, however, the microplasticity-effect (a seemingly smaller Young's modulus) is more pronounced because a small bow out of a favorably oriented FR source at the beginning of the simulations is more likely.} {\rev When GB effects and bulk hardening are neglected in the GP model at hand, the defect energy exclusively controls the {\redsw overall} rate of work hardening. {\rtwo The plastic defect parameter is fitted to a value of \f{K_{\rm G}=18\times10^{-6}}~N}.} \rrevC{\figref{fig:lc0elplel_gbhard} shows that} the used quadratic form of the defect energy leads to an overestimation of plastic strain in the center of the grain. {\rev Close to the GBs, the plastic strain is underestimated as are the strain gradients. The overall shapes of the plastic strain profiles from both simulation approaches differ more pronounced \rrevC{from each other}, compared to the cases considered above}.
\FF{htbp}
\includegraphics[width=\linewidth]{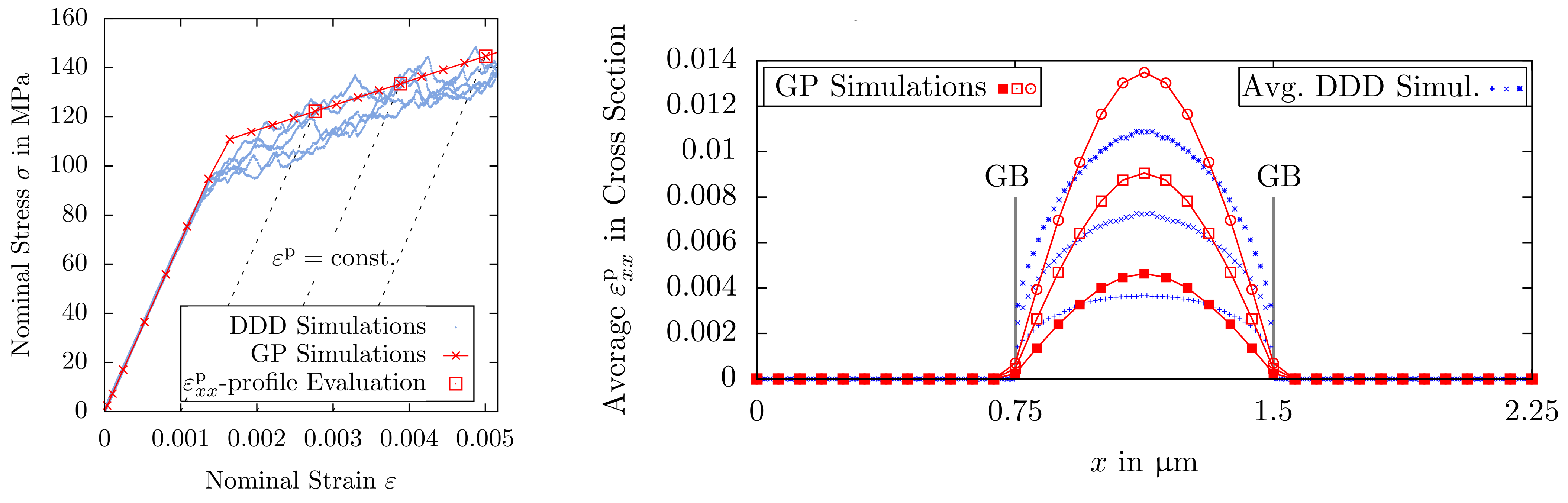}
\FC{LC0E: Stress-strain curves of DDD and GP simulations (left). Distribution of cross-section averaged plastic strain along the loading direction of DDD and GP simulations (right, values are obtained at fixed overall plastic strain values, cf.\ left). Boundary conditions with free lateral contraction, all three grains in $\langle 100 \rangle$-orientation. Bounding grains behave elastic (i.e., $\tau_{0}^{\rm C}\rightarrow\infty$ in the GP model of these). GB hardening is neglected in the GP model, as are the GB yield strength and the Voce hardening}
\label{fig:lc0elplel_gbhard}
\Ff

\section{Discussion of Results}
\subsection{{\rev Grain Boundary Yielding} in the Gradient Model}
From the above results, it can be seen that GB yielding is mandatory in the GP model, if the plastic strain profiles obtained from DDD simulations with impenetrable GBs should {\rev be captured.} The use of common microhard conditions for the GBs, i.e., an infinitely high GB yield strength~\f{\Xi_0^{\rm C}} in the model at hand, is not appropriate. {\blue This is due to the fact that the DDD profiles show that plastic strain, and thus, plastic slip, is present in the immediate vicinity of the GBs.} {\orange Microhard conditions, however, have been used in the GP model of \cit{aifantis2009discrete}, where the DDD simulation setup and framework is similar to the one in the present work. The evaluation of the plastic strain profiles \rrevC{differs from the present work}: there, the DDD strain profiles are obtained by integrating the dislocation density tensor with an integration constant, yielding vanishing plastic strain at the GB. In the present work, however, the strain profiles are obtained by slicing the sample and using the {\rev averaging} described in~\secref{sec:ddd_averaging}, leading to non-zero values of the plastic strain~\f{\varepsilon^{\rm p}_{xx}} in the immediate vicinity of the GBs. \roneC{Due to the volumetric averaging, \f{\varepsilon^{\rm p}_{xx}} is evaluated in the slices adjacent to the GBs \rrevC{but} not in the GB planes themselves.}
% Since this is a volumetric procedure, the plastic strain value at the GB itself is not evaluated in the DDD framework.}
\subsection{{Grain Boundary Hardening in the Gradient Model}}
 In order to account for the pile-ups of the DDD results, an additional hardening relation {\rev is} introduced in the GP model. {\rev It is associated to the micromorphic field variable~\f{\zeta} and, consequently, to the equivalent plastic strain~\f{\gaq}, at the GB itself.} Thus, the GB yield strength increases with rising~\f{\gaq}. The DDD plastic strain profiles and their evolution can not be captured, if, instead of GB hardening, solely a {\rev bulk hardening relation} of Voce type is used in the GP model. {\rev This is especially apparent close to the GBs,  cf.~\figref{fig:nlc35_gbhard}.} The DDD results can, furthermore, be {\rev captured} notably well close to the boundaries with this procedure, for the investigated cases at hand. Consequently, this indicates that {\rev hardening associated to the boundaries as well is necessary in the GP model} to account for the BCs of the DDD simulations.\\
{\green In addition, the use of a finite GB strength~\f{\Xi_0^{\rm C}}, in combination with a sufficiently large GB hardening parameter~\f{K_{\rm H}\rightarrow\infty}, reproduces GB behavior similar to microhard conditions (cf.~\figref{fig:k_h_study}). However, in this combination, the model behavior allows for some plasticity to occur at the GB, at first. Subsequently, by the high GB hardening contribution to the GB yield strength, plasticity is prevented from proceeding further. The deviations of this approach from ideal microhard conditions (i.e., \f{\Xi_0^{\rm C}\rightarrow\infty}) are small for the overall mechanical responses {\rev and for the plastic strain profiles} .
%  Nevertheless, it is remarked that slight differences can occur in the associated plastic strain profiles.}
\subsection{Dislocation Interaction across Grain Boundaries}
It is challenging to incorporate dislocation interaction across GBs in GP simulations with the same physical details as in DDD simulations. {\rev With the used GP model in this work, identical plastic strain profiles are obtained for two different crystal orientations of the central grain (cf.\ \figref{fig:nlc35_gbhard} and \figref{fig:nlc5_gbhard}).} 
% The GP simulations {\blue considering Voce hardening, as well as considering GB hardening, respectively, each yield almost identical plastic strain profiles in spite of the different crystal orientations of the central grain (cf.\ NLC5G and NLC35G).}
Contrarily, the DDD simulations show significantly different plastic strain profiles due to the {\rev orientation dependent} dislocation interaction across GBs. {\blue While no dislocations are allowed to pass the {\rev GBs} in the DDD setting, the stress fields of dislocation pile-ups on adjacent sides of the GBs interact nevertheless. This influences the plastic strain gradients close to the GBs, depending on the misorientation between the grains. A high misorientation, resulting in a large mismatch of the respective stress fields, {\rev induces a parabolic distribution of plastic strain (cf.\ \figref{fig:nlc35_gbhard})}. In this case, the interaction between dislocations across the GBs is weaker than \rrevC{it is} for lower misorientations. Thus, the central grain becomes the main carrier of plasticity in the DDD simulations.} Regardless of the phenomenological approach to GB strengthening, the plastic strain profile in the central grain is reasonably captured by the GP model. Contrarily, a low misorientation leads to a more homogeneous distribution of the plastic strain in the grains with sharper ``cusps'' towards the GBs (cf.~\figref{fig:nlc5_gbhard}). In this case, the DDD and GP profiles are in rather good agreement in the two bounding grains. Thus, further research regarding the appropriate modeling of the dislocation interaction across the GBs in the continuum model {\rev is} necessary.
\subsection{Evolution of Plastic Strain Gradients}\label{sec:plastsg}
{\rev The gradients of the plastic strain, close to the GBs in the central grain of the DDD results, are reasonably captured by the first GP strain profile at~\f{\varepsilon_{\rm p}=0.001} (cf.\ \figref{fig:nlc35_gbhard} and \figref{fig:nlc5_gbhard})}. \roneC{Furthermore, it can be observed that the GP model with GB hardening does capture the evolution of plastic strain values close to the GB, but not the evolution of the gradients, there. One could argue that this might stem from the employed GP model with~\f{\gaq}, \f{\zeta}, and its gradient, in general. The investigated special case LC0E, however, \rrevC{gives} indications that the differences in the gradients between GP and DDD results could mainly be caused by the convenient choice of a quadratic defect energy. A more general form of the defect energy that could also be formulated with respect to~\f{\nabla\zeta} (e.g., using a power law approach~\cip{bardella2010size}) could give better agreement in the gradients. This could also be obtained by applying more physically enriched theories, e.g., \cip{bardella2013latent,gurtin2014gradient,bayley2006comparison}, allowing to consider dislocation densities and a physically more sound defect energy definition, respectively.} {\rthr One might further argue that the evaluation of strain gradients is dependent on \rrevC{the} discretization, especially in the DDD setting. It can, however, be shown that the plastic strain values close to the GB encounter only minor differences upon further discretization refinement in the DDD simulations. This is due to the fact that the discretization resolution is chosen higher than the dislocation spacing in the pile-ups (cf.~\secref{sec:ddd_averaging}).}{\rev\ The GP strain profiles show more pronounced hyperbolic distributions than the DDD strain profiles do, independent of the chosen misorientations. \roneC{The DDD strain profiles (and gradients), thus, are obviously sensitive to the misorientation and show the role of dislocation interaction across the GBs, a feature, the GP model at hand cannot account for. This \rrevC{shortcoming} could be removed by using, e.g., one of the above-mentioned approaches of \cip{bardella2013latent,gurtin2014gradient,bayley2006comparison}.}
\subsection{Assumptions in the Gradient Plasticity Model}
\label{subsec:assumptions}
 \roneC{In the GP model, an equivalent plastic strain is taken into account. Its micromorphic counterpart~\f{\zeta} is assumed to be continuous across the GBs while the gradient~\f{\nabla\zeta}} is not. It is remarked that employing continuity of, e.g., plastic strain, but allowing for jumps in its gradient, has shown good agreement with the size effects observed in experiments, in a one-dimensional case (cf.\ \cip{aifantis2006interfaces}, and the discussion in the introduction of the work at hand). Furthermore, the equivalent plastic strain of the present work is the (directionally independent) sum of the slip contributions of all slip systems. Therefore, it is not a quantity from which conclusions on the behavior of individual slips should be drawn, but, rather, on an overall response of these. Consequently, the choice of continuity for~\f{\zeta} across GBs is not expected to be in general transferable to statements about the continuity of slips on individual slip systems across GBs. These are known to be often discontinuous across GBs due to different operating mechanisms during slip transfer (e.g., \cip{lee1990tem}). Within the context of an overall description of plastic slip, it might also be noteworthy that the calculations of an effective plastic strain in the experimental work of \cip{abuzaid2012slip} lead to a continuous appearance of distributions of this quantity across many GBs. Furthermore, the DDD simulations in the work at hand show similar values of the plastic strain on the adjacent sides of the GBs in $x$-direction. As there is a formal connection between~\f{\gamma_{\rm eq}} (and thus \f{\zeta}) and the components of~\f{\feps^{\rm p}}, this finding might indicate that the continuity assumption on \f{\zeta} could be justifiable for certain cases.}\\  {The internal length scale in the GP model can be deduced from the defect energy parameter~\f{K_{\rm G}}. The used value of \f{K_{\rm G}} (cf.~\secref{sec:parameters}) yields an internal length of \f{l=\sqrt{K_{\rm G}/E}\approx36}~nm, considering a Young's modulus of 65~GPa. Remarkably, this result is of the same order of magnitude as the mean dislocation spacing in the pile-ups at the GBs of the DDD simulations (cf.~\secref{sec:ddd_averaging}). {\blue Additionally, it is not expected that~\f{K_{\rm G}} (and thus, the internal length scale) being a constant should be generally valid.}}
\section{Conclusions}
For the modeling of pile-ups, as they are observed in discrete dislocation dynamics simulations with impenetrable grain boundaries, grain boundary yielding is used in the gradient plasticity model \rrevC{of} this work. {\rev The increasingly entangling dislocations at the grain boundaries are modeled by using grain boundary hardening.}. Thereby, the overall mechanical response of discrete dislocation dynamics simulations as well as the local response in form of plastic strain profiles is reasonably captured for certain cases. The comparison of this approach to a previous model without grain boundary hardening, but with an isotropic Voce hardening relation instead, shows inferior agreement. Without the grain boundary hardening, the evolution of plastic strain values, observed in the immediate vicinity of grain boundaries in discrete dislocation dynamics simulations, can not be captured by the continuum model. Microhard grain boundary conditions, commonly used in continuum models in the literature, are not sufficient to account for the observed behavior at the impenetrable grain boundaries of discrete dislocation dynamics simulations. Contrary to these common {\rev conditions}, {\rev rather} a finite grain boundary yield strength {\rev needs to be} used. The plastic strain profiles obtained from the continuum model are in agreement with the discrete dislocation dynamics results {\rev in the immediate vicinity of the grain boundary}. \rroneC{It could also be argued that physically more advanced theories (by incorporating higher-order terms) might account for the sharp gradients close to grain boundaries - in spite of using microhard conditions.} \rrthrC{The present GP model, however, allows for extended fully threedimensional simulations in manageable computational times, a benefit often lost when using physically enriched theories with substantially increased degrees of freedom.} \rthrC{Differences between the results of the continuum and the discrete modeling approach indicate that additional research is necessary related to accounting for crystal orientation dependent dislocation interaction mechanisms across grain boundaries in the continuum model. \rrtwoC{This also includes the possible recalibration of the model parameters for different orientations and loading conditions as well as further case studies}.} Additionally, the question of {\rev an enhanced defect energy approach} is raised, as well as a suitable combination of hardening approaches considering both, geometrically necessary dislocations and statistically stored dislocations. \rthrC{Further investigations could also consider elastic anisotropy and different boundary conditions.}
{\appendix
\section{Parameter Study of Grain Boundary Hardening}
\setcounter{figure}{0}
{\centering
\FF{htbp}\includegraphics[width=0.4\linewidth]{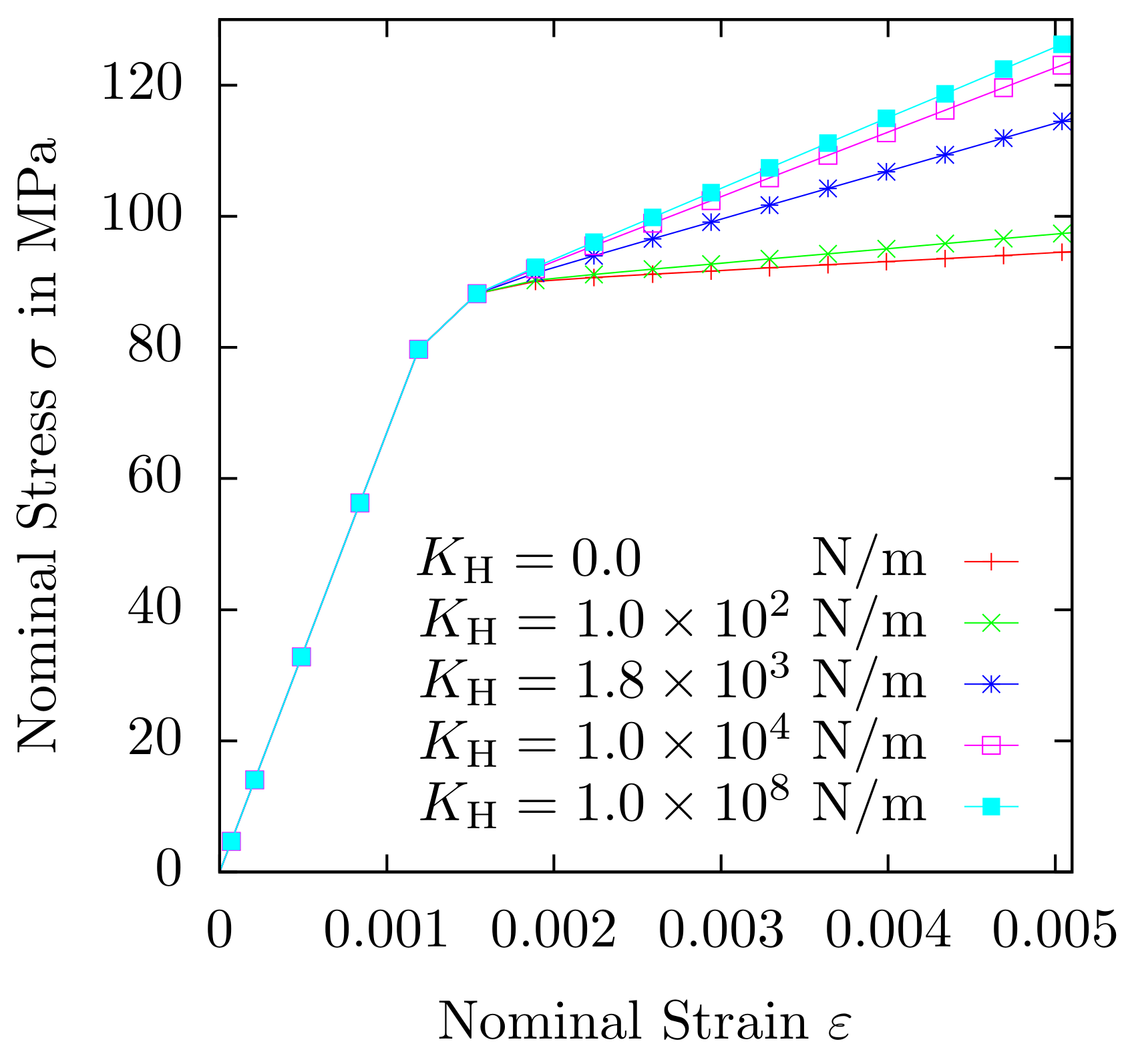}
\caption{\green Parameter study of grain boundary hardening parameter~\f{K_{\rm H}}. Case NLC35G, only~\f{K_{\rm H}} is varied as shown above. All other parameters can be found in~\tabref{tab:setups} and \secref{sec:parameters}, respectively.}
\label{fig:k_h_study}
\Ff}}
% \newpage
\rrevC{
% {\bf Data accessibility.} This work does not report any data.\\
{\bf Competing interests.} The authors declare that there are no competing interests.\\
{\bf Authors' contributions.} E.B., M.S.\ carried out the work and drafted the manuscript. S.W.\ supported conceptual and theoretical design. D.W.\ and T.B.\ coordinated and reviewed work / manuscript. All authors gave final approval of the manuscript.\\
{\bf Funding statement.} The authors acknowledge the support rendered by the German Research Foundation (DFG) under Grants BO1466/5-1 and WE3544/5-1. The funded projects "Dislocation based Gradient Plasticity Theory" and "Discrete Dislocation Dynamics" are part of the DFG Research Group 1650 "Dislocation based Plasticity".\\
DDD computations were performed on bwUniCluster funded by Ministry of Science, Research and Arts / Universities of the State of Baden-W\"urttemberg, Germany, framework program bwHPC, on the (DFG funded) IC2, and the HC3 at the SCC at KIT.}
% \input{07_appendix_zeta}
% \newpage
% {\bf References}\\[1ex]
\bibliographystyle{elsarticle-harv}
\bibliography{lit_abb}
\end{document}